\documentclass[aps,pra,twocolumn,groupedaddress,amssymb]{revtex4-1}
\usepackage{microtype}
\usepackage{latexsym}
\usepackage{amsbsy}
\usepackage{amsmath}
\usepackage{graphicx}
\usepackage{color}

\usepackage{color}

\begin{document}
\date{\today}

\title{Controlled creation and stability of $k\pi$-skyrmions on a discrete lattice}

\author{Julian~Hagemeister}
\author{Ansgar~Siemens}
\author{Levente~R\'ozsa}
\author{Elena~Y.~Vedmedenko}
\email{vedmeden@physnet.uni-hamburg.de}
\author{Roland~Wiesendanger}

\affiliation{Department of Physics, University of Hamburg, D-20355 Hamburg, Germany}

\date{\today}

\begin{abstract}
\noindent We determine sizes and activation energies of $k\pi$-skyrmions on a discrete lattice using the Landau-Lifshitz-Gilbert equation and the geodesic nudged elastic band method. The employed atomic material parameters are based on the skyrmionic material system Pd/Fe/Ir(111). We find that the critical magnetic fields for collapse of the $2\pi$-skyrmion and $3\pi$-skyrmion are very close to each other and considerably lower than the critical field of the $1\pi$-skyrmion. The activation energy protecting the structures does not strictly decrease with increasing $k$ as it can be larger for the $3\pi$-skyrmion than for the $2\pi$-skyrmion depending on the applied magnetic field. Furthermore, we propose a method of switching the skyrmion order $k$ by a reversion of the magnetic field direction in samples of finite size. 
\end{abstract}

\maketitle

\section{Introduction}

Magnetic skyrmions were originally predicted theoretically as localized noncollinear textures in crystals lacking centrosymmetry\cite{Bogdanov1989}. Later it was realized that magnetic skyrmions represent a member of a larger family of rotationally symmetric magnetic solitons\cite{Bogdanov1999}. In these structures, the magnetization direction rotates by an angle of $k\pi$ between the center and the ferromagnetic background. The focus of most research activities has been the structure in which the angle of magnetization rotation is $1\pi$, generally simply referred to as a magnetic skyrmion within the community. Experimentally, such skyrmions have been observed in chiral bulk helimagnets\cite{Muhlbauer2009,Pappas2009,Yu2010,Yu12011,Seki2012,Tonomura2012,Milde2013,Park2014} as well as in ultrathin magnetic film systems\cite{Romming2013}. They have been heavily investigated both theoretically and experimentally regarding their thermal stability \cite{Hagemeister2015} and their response to a spin-polarized electric current\cite{Jonietz2010,Zang2011,Yu2012,Sampaio2013,Iwasaki2013,Iwasaki2013a,Iwasaki2014,Schutte2014b,Knoester2014,Woo2016} in order to realize skyrmionic racetrack devices\cite{Fert2013} at room temperature in the future.

In comparison to the $1\pi$-skyrmions, $k\pi$-skyrmion structures with $k>1$ have received less attention both from the experimental and the theoretical side up to now. A reason for this may be that they are difficult to create in experiments from a technical point of view; the observation of structures with $2\pi$ magnetization rotation has been restricted to materials with nonnegligible dipolar interaction and a vanishing Dzyaloshinskii--Moriya interaction so far\cite{Finazzi2013,Streubel2015}.
Previous theoretical investigations dealt with magnetization profiles and the size of individual $k\pi$-skyrmions as a function of external magnetic field in systems of infinite size\cite{Bogdanov1999}. 

For a quasi two-dimensional system with a continuous magnetization field of unit length $\textbf{S}$, the topological charge is given as a surface integral over the area $A$\cite{Rozsa2017},
\begin{align}
Q = \frac{1}{4 \pi}\int_A{\textbf{S}\left(\frac{\partial \textbf{S}}{\partial x}\times\frac{\partial \textbf{S}}{\partial y} \right)dxdy}=-\frac{1}{2}\left[\cos\Theta\left(r\right)\right]_{0}^{\infty}m.
\label{eqn:topo}
\end{align} 

For $k\pi$-skyrmions the vorticity is $m=1$ and the polar angle $\Theta\left(r\right)$ changes by $k\pi$, meaning that the topological charge is $Q=0$ for even $k$ and $\left|Q\right|=1$ for odd $k$.
Studies of the propagation of $2\pi$-skyrmions caused by an applied field gradient\cite{Komineas2015} and spin-polarized electric currents\cite{Zhang2016} revealed major differences compared to the $1\pi$-skyrmion state. While $1\pi$-skyrmions of topological charge $Q=1$ experience a deflection due to the skyrmion Hall effect\cite{Jiang2017}, $2\pi$-skyrmions are expected to not show this effect since their topological charge is $Q=0$. 

The injection of perpendicular spin-polarized currents\cite{Zhang2016} and the use of laser beams with orbital angular momentum\cite{Fujita2017,Fujita2017a} have been proposed as means to create $k\pi$-skyrmionic states.
Moreover, there are several studies dealing with the stability of $k\pi$-skyrmions in disk-shaped samples of finite size\cite{Rohart2013,Leonov2014,Mulkers2016}. For such systems, it was found that the transition between different states is governed by a Bloch point propagation which can be linked to a breathing-type excitation of the skyrmionic structure\cite{Beg2015,Liu2015}. 
However, a procedure to reliably switch between different $k\pi$-skyrmions is still lacking.

Here, we discuss the size dependence of $k\pi$-skyrmions with $k\in \{1,2,3\}$ in a spin lattice model on the external magnetic field and compare their minimal size to the previously determined minimal size of the $1\pi$-skyrmion in Ref.~\cite{Siemens2016}. Moreover, we calculate the energy barriers associated with $k\pi$-skyrmions with $k\in \{1,2,3\}$ in infinite systems in order to determine the feasibility of stabilizing them in experiments. We also propose a simple method of reliably switching between skyrmionic states of different order in finite samples by nonadiabatic reversals of the external magnetic field.

\section{Simulation methods}

We describe the behavior of classical Heisenberg spins $\{\textbf{S}_{i}\}$ with $|\textbf{S}_{i}| = 1$ on a two-dimensional triangular lattice with lattice constant $a$ using the Hamiltonian 
\begin{align}
H = &-J \sum_{<i,j>}\textbf{S}_{i} \cdot \textbf{S}_{j} - \sum_{<i,j>} \textbf{D}_{i,j} \cdot \left( \textbf{S}_{i} \times \textbf{S}_{j} \right) \nonumber\\
&- K \sum_{i} S_{i,z}^2 -\mu \sum_{i}\textbf{S}_{i} \cdot \textbf{B}, 
\label{eqn:hamiltonian}
\end{align} 
being standard for skyrmionic systems. Therein, $J$ is an effective exchange interaction coefficient, $\textbf{D}_{i,j}$ is the Dzyaloshinskii--Moriya vector, $K$ the on-site anisotropy energy parameter, $\mu$ the atomic magnetic moment, and $\textbf{B}$ the external magnetic field. The summations in Eq.~(\ref{eqn:hamiltonian}) are understood over all nearest-neighbor pairs for the exchange and Dzyaloshinskii--Moriya interactions. We use interaction parameters for Pd/Fe/Ir(111) which is the first ultrathin magnetic film system found to exhibit individual skyrmions\cite{Romming2013}. The micromagnetic parameters derived experimentally in Ref.~\cite{Romming2015} are adapted to the discrete lattice, with $J = 5.72\,$meV, $D=|\textbf{D}_{i,j}|=1.52\,$meV and $K=0.4\,$meV. For a comparison of discrete and continuum model verifying the correctness of the discrete energy parameters see Appendix \ref{app}. An effective atomic magnetic moment of $\mu=3\,\mu_{\mathrm{B}}$ is used in Eq.~(\ref{eqn:hamiltonian}) based on density functional theory calculations\cite{Dupe2014}, with $2.7\,\mu_{\mathrm{B}}$ stemming from the Fe atoms and $0.3\,\mu_{\mathrm{B}}$ from the Pd atoms.

For obtaining the equilibrium spin states, as well as for performing the dynamical calculations, the stochastic Landau-Lifshitz-Gilbert (LLG) equation\cite{Landau,Gilbert} including thermal fluctuations was used,
\begin{align}
\partial_t \textbf{S}_{i} = &-\gamma'\textbf{S}_{i}\times(\textbf{B}_i^{\mathrm{eff}}+\textbf{B}_i^{\mathrm{th}})\nonumber\\
&-\alpha\gamma'\textbf{S}_{i}\times(\textbf{S}_{i}\times(\textbf{B}_i^{\mathrm{eff}}+\textbf{B}_i^{\mathrm{th}}))
\label{eqn:llg}
\end{align}  
with the gyromagnetic ratio $\gamma=ge/2m$ ($g,e,$ and $m$ are the electron $g$ factor, charge, and mass, respectively), the Gilbert damping parameter $\alpha$, and $\gamma'=\gamma/(1+\alpha^2)$. The effective field $\textbf{B}_i^{\mathrm{eff}}$ is given by
\begin{align}
\textbf{B}_i^{\mathrm{eff}} &= -\frac{1}{\mu}\frac{\partial H}{\partial \textbf{S}_{i}},
\label{eqn:fields}
\end{align} 
while the variance of the fluctuating thermal field $\textbf{B}_i^{\mathrm{th}}$ is proportional to the temperature $T$.

For the calculation of energy barriers, we used the \textit{Spirit} simulation code\cite{Spirit}
which offers the possibility to relax a system to a local energy minimum with the LLG equation and to perform calculations using the geodesic nudged elastic band (GNEB) method\cite{Bessarab2015}, which has been successfully applied to the investigation of $1\pi$-skyrmions in the past\cite{Lobanov2016,Stosic2017}. The GNEB method finds a minimum energy path between two distinct magnetic states as well as the saddle point which can be identified as the minimal-energy transition state between the end points. The one-dimensional reaction coordinate connecting the states is defined as
\begin{align}
\Delta R_{ij} = \sqrt{\sum_k\left(\alpha^k_{i,j}\right)^2}\label{eqn:rc}
\end{align}
measuring the difference between two spin configurations $i$ and $j$ of the system.
Therein, $\alpha^k_{i,j}$ is the angle between a spin at site $k$ in configuration $i$ and the spin at the same site in configuration $j$.

\section{Results}

\subsection{Equilibrium size of $k\pi$-skyrmions}

\begin{figure}[tb]
\includegraphics[width=\columnwidth]{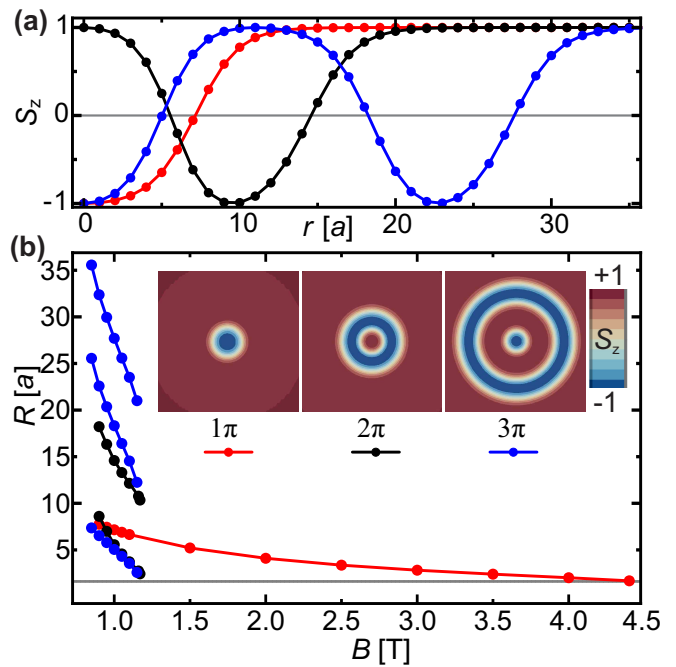}
		\caption{(a) Magnetization profiles of $k\pi$-skyrmions as a function of the distance $r$ to the center of the respective structure at $B=1\,$T. (b) Radii of the rings where $S_\mathrm{z}=0$ for the $k\pi$-skyrmionic states as a function of the magnetic field. The number of rings in which the magnetization lies within the sample plane is equal to the skyrmion order $k$. The inset shows color-map plots of the z component for the $k\pi$-skyrmionic states at $B=1\,$T.}
		\label{graphic:fig1}
\end{figure}
In the following, we discuss the size of $k\pi$-skyrmions ($k\in\{1,2,3\}$) as a function of external field, as well as the critical field value where they collapse on the lattice. See Fig.~\ref{graphic:fig1}(a) and the inset of Fig.~\ref{graphic:fig1}(b) for exemplary magnetization profiles and color-map plots of the out-of-plane magnetization component $S_\mathrm{z}$ for these structures. Following the definition of previous investigations\cite{Romming2015}, we determine the size of $k\pi$-skyrmions by the radii of the rings in which the magnetization direction locally lies within the plane of the two-dimensional spin lattice. The number of such rings is equal to the skyrmion order $k$ and hence, the ring with the largest diameter provides a good estimate of the total size of the respective structure.

The radii of the rings with $S_\mathrm{z} = 0$ are shown in Fig.~\ref{graphic:fig1}(b) as a function of the external magnetic field $B$. For the calculations, we used a hexagonal-shaped sample with 10621 spins and kept the spins at the edge fixed parallel to the magnetic field. The sample diameter was at least twice as large as the diameter of the considered $k\pi$-skyrmions, ensuring that edge effects played a minor role in the simulations.
For all investigated magnetic fields, the size of the skyrmionic structures increases with the order $k$. At $B=1\,$T in Fig.~\ref{graphic:fig1}(a), the $2\pi$-state is about twice (2.04) as large as the $1\pi$-skyrmion, while the $3\pi$-skyrmion is by about a factor of four (3.9) larger than the $1\pi$-skyrmion. However, note that these ratios also depend on the magnetic field.
The size of $k\pi$-skyrmions decreases with an increasing external magnetic field up to a finite critical field at which the $k\pi$-skyrmion becomes unstable. This critical field is found to decrease with increasing skyrmion order $k$; we identified the values $B_{c,1\pi}=(4.495\pm 0.005)\,$T,  $B_{c,2\pi}=(1.175 \pm 0.005)\,$T and $B_{c,3\pi}=(1.155\pm 0.005)\,$T for the $1\pi$-, $2\pi$-, and $3\pi$-skyrmion states, respectively. Interestingly, the critical fields of the $2\pi$- and $3\pi$-skyrmions are very close to each other and much smaller than that of the $1\pi$-skyrmion.
At the critical field, the compression of a $k\pi$-skyrmion structure due to the external magnetic field leads to the annihilation of the center ring, which effectively transforms the structure into a $(k-1)\pi$-skyrmion. 

\subsection{Energy barriers}
\begin{figure}[tb]
\includegraphics[width=\columnwidth]{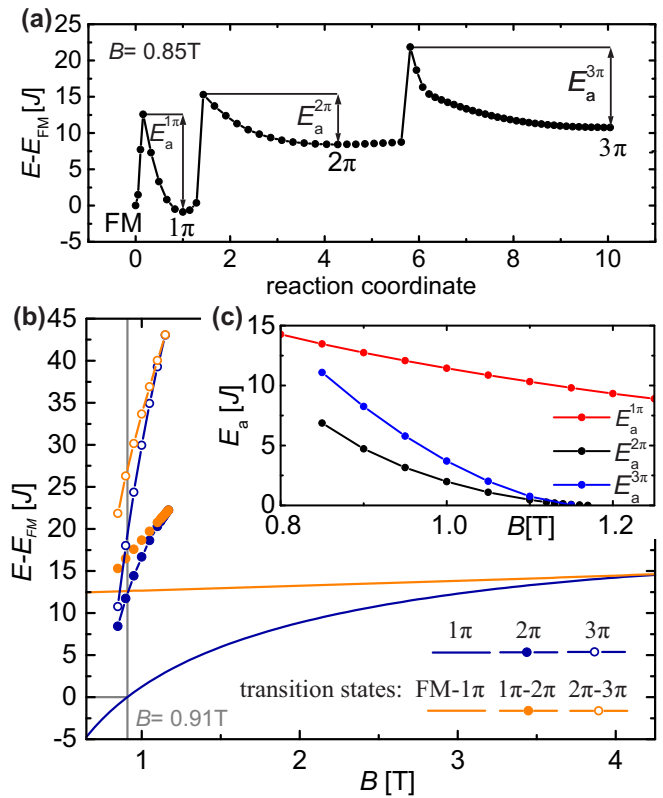}
		\caption{(a) Minimum energy path between $k\pi$-states derived using the GNEB method for the magnetic field $B = 0.85\,$T. (b) The energies of the $1\pi$-, $2\pi$- and $3\pi$-skyrmions and of the three transition states (FM-$1\pi$,$1\pi$-$2\pi$,$2\pi$-$3\pi$) with respect to the energy of the ferromagnetic configuration as a function of the magnetic field, given in units of the nearest-neighbor exchange interaction $J$. (c) The activation energies for a transition from a $k\pi$-state to a $(k-1)\pi$-state.}
		\label{graphic:fig2}
\end{figure}

In order to achieve a deeper understanding of the stability of $k\pi$-skyrmionic states, we calculated the energy barriers with the GNEB method.
In order to minimize boundary effects, we chose samples with periodic boundary conditions and diameters $d$ which are large compared to the radius $R$ of the considered $k\pi$-skyrmionic state $(d>4R)$. Figure~\ref{graphic:fig2}(a) shows the minimum energy path from the ferromagnetic (FM) state through the $1\pi$- and $2\pi$- to the $3\pi$-skyrmion at $B=0.85\,$T. The reaction coordinate from Eq.~(\ref{eqn:rc}) is normalized such that the distance between the ferromagnetic state and $1\pi$-skyrmion is $1$. Note that the distance between two configurations in principle depends on the number of interior configurations due to the discretization of the minimum energy path; however, we confirmed that the relative distances between the different magnetic states (FM, $1\pi$, $2\pi$, $3\pi$) in Fig.~\ref{graphic:fig2}(a) are relatively robust with respect to the number of configurations used in the GNEB calculations. Hence, it can be concluded that the distance in phase space from the $2\pi$-state to a neighboring transition state is much larger than the distance from the $1\pi$-state to a transition state for this particular magnetic field, but the distances decrease as the critical field is approached. 

Furthermore, we investigated energies of the local energy minima and the transition states as a function of the external magnetic field, see Fig.~\ref{graphic:fig2}(b). The energies of $1\pi$-, $2\pi$- and $3\pi$-skyrmions increase with respect to the energy of the ferromagnetic state as a function of the magnetic field. This is expected because the skyrmions contain spins aligned oppositely to the external field leading to a loss in Zeeman energy, in agreement with typical phase diagrams\cite{Bogdanov1994} describing skyrmionic systems which show the ferromagnetic state as the ground state for large magnetic fields. The energy of a single skyrmion in a ferromagnetic background is equal to the energy of the ferromagnetic state at $B=0.91\,$T, which should be equal to the transition field from the skyrmion lattice phase to the ferromagnetic phase. The energies of the $2\pi$- and $3\pi$-states are much more sensitive to changes in the magnetic field than the energy of the $1\pi$-skyrmion state.

The activation energies, defined as the energy difference between the local energy minimum and the transition state (see also Fig.~\ref{graphic:fig2}(a)), are displayed in Fig.~\ref{graphic:fig2}(c). While the $1\pi$-skyrmion is the most difficult to collapse at low fields, surprisingly we find that the activation energy $E_a^{3\pi}$ protecting the $3\pi$-state against external perturbations is by about a factor of $1.6-1.8$ larger than that of the $2\pi$-state for magnetic fields smaller than $B<1.1\,$T. However, the activation energies intersect between $B=1.1\,$T and $B=1.15\,$T with $E_a^{3\pi}<E_a^{2\pi}$ at $B=1.15\,$T. The $3\pi$-skyrmion collapses at a slightly lower field value ($B_{c,3\pi}=1.155\,$T) than the $2\pi$-skyrmion ($B_{c,2\pi}=1.175\,$T) where the energy barrier protecting it becomes zero, in agreement with the spin dynamics simulations above.

\subsection{Minimal sizes}
On a discrete lattice, a skyrmion cannot shrink to arbitrarily small sizes, instead there exists a finite minimal size which is approached close to the critical magnetic field. This is indicated by the gray line in Fig.~\ref{graphic:fig1}(b) for the $1\pi$-skyrmion, corresponding to the smallest skyrmion size close to the critical field where the skyrmion could still be stabilized. A similar method for approximately determining the minimal size was used in a previous study\cite{Siemens2016}.

The GNEB calculations also offer the opportunity to investigate the minimal sizes of the $k\pi$-skyrmionic structures on a discrete lattice. However, in the GNEB method it is possible to give a more rigid definition of the minimal size of a $k\pi$-skyrmionic structure, since it provides direct access to the spin configurations of the transition states as a function of the external magnetic field. From the minimum energy path, it is known that with an infinitesimal decrease or increase in the diameter of the transition state structure between the $(k-1)\pi$ and $k\pi$ state, the system will converge either to the $(k-1)\pi$ or $k\pi$ state. Hence, the transition state defines the minimal size of the $k\pi$-skyrmion for a given external magnetic field.

\begin{figure}[tb]
\includegraphics[width=\columnwidth]{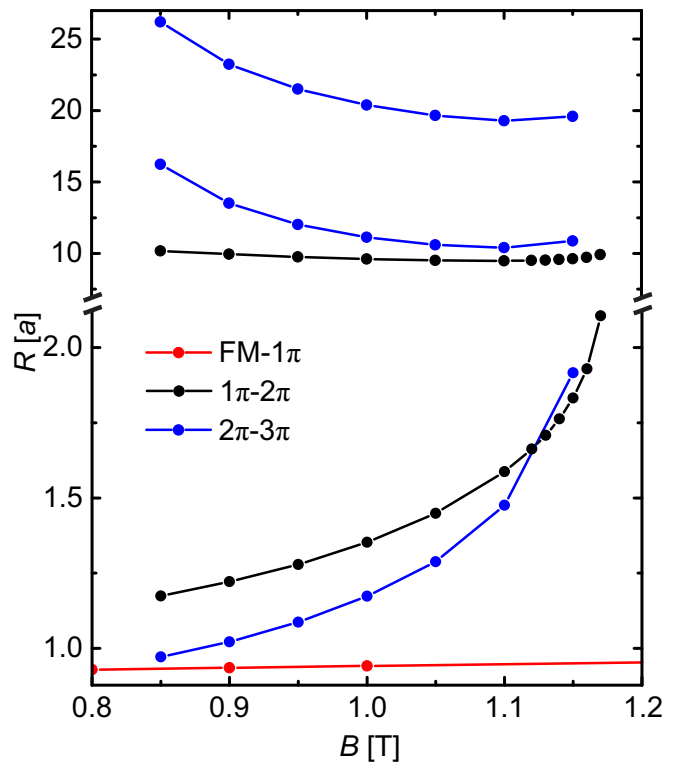}
		\caption{Radii of the rings where $S_\mathrm{z} = 0$, for the transition states obtained with the GNEB method as a function of the external magnetic field. Note the change in the vertical scale between the first and second rings.}
		\label{graphic:fig3}
\end{figure}
Figure~\ref{graphic:fig3} shows the minimal sizes of $k\pi$-skyrmions derived from the transition states as a function of the external magnetic field, where the ring sizes were obtained the same way as in Fig.~\ref{graphic:fig1}. The minimal size of the $1\pi$-skyrmion increases with increasing magnetic field. This can be explained by the fact that the skyrmion becomes energetically less favorable for larger magnetic fields; therefore, the size to which it has to be compressed for annihilation becomes larger with an increasing magnetic field. A similar behavior can be observed for the radii of the innermost rings of the $2\pi$- and $3\pi$-skyrmions. On the other hand, it is interesting to note that the radii of the outer rings of the $2\pi$- and $3\pi$-skyrmions exhibit a minimum below the critical field.

\subsection{Creation of $k\pi$-skyrmions}

\begin{figure}[bt]
\includegraphics[width=\columnwidth]{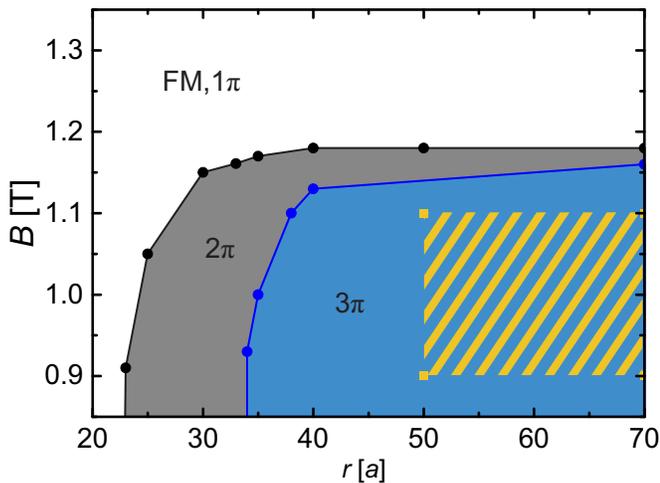}
		\caption{Phase diagram showing regions of metastability of FM, $1\pi$-, $2\pi$-, and $3\pi$-states as a function of disk radius and external magnetic field. The FM and $1\pi$-states are stable in the whole depicted area, with the $2\pi$- and $3\pi$-skyrmions additionally stabilized inside the black and blue regions, respectively. The region marked with orange lines indicates the system sizes and magnetic fields for which the transition from a $(k\pi)$ to a $(k+1)\pi$ state due to the reversal of the magnetic field direction was investigated.
}
		\label{graphic:fig4}
\end{figure}

In the following, we propose a method to reliably change between skyrmionic states of different order $k$ in disk-shaped magnetic samples with a radius $r$ and open boundary conditions. Considering the field dependence of the sizes of skyrmionic structures discussed in the previous sections, first we had to determine a range of fields and radii where the $1\pi$-, $2\pi$- and $3\pi$-skyrmions may all be stabilized inside the disk. A corresponding phase diagram determined using LLG simulations is shown in Fig.~\ref{graphic:fig4}. Note that small disk sizes cannot accommodate $k\pi$-skyrmions consisting of many rings, while large magnetic fields lead to a collapse of the inner rings. Consequently, we performed the following dynamical simulations in the $0.9\,\textrm{T}\le B\le 1.1\,\textrm{T}$, $50a\le r\le70 a$ region denoted by dashed orange lines in Fig.~\ref{graphic:fig4}.

\begin{figure}[tb]
\includegraphics[width=\columnwidth]{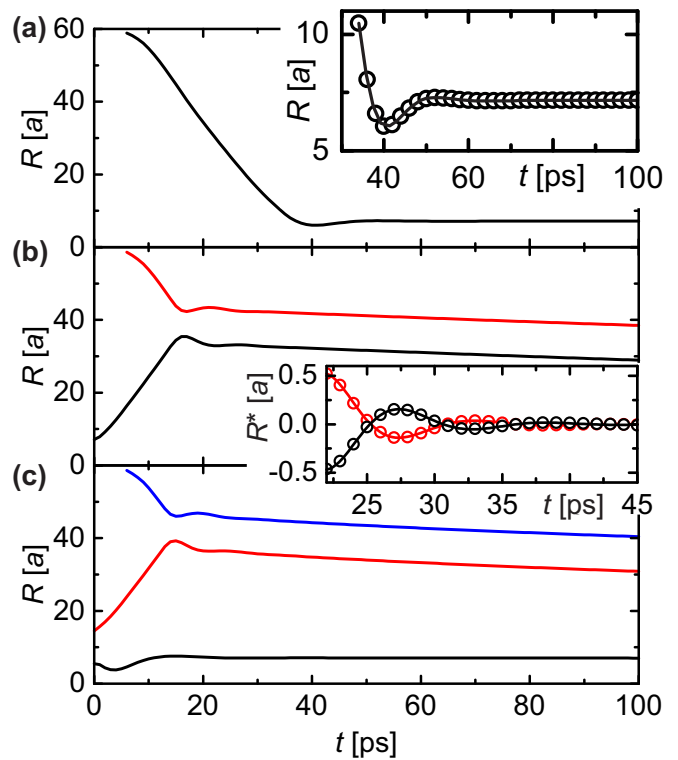}
		\caption{The dynamic process of the creation of $k\pi$-skyrmions. The initial configuration is a relaxed ferromagnetic disk with $r=60\,a$ at $B=1\,$T. The direction of the external field is instantaneously reversed at $t=0\,$ps and a skyrmion is formed via the sample edge as a function of time in (a). The inset shows a magnification of the oscillation of the skyrmion radius about $35\,$ps after the field reversal, fitted by damped oscillatory function in Eq.~(\ref{eqn:1}) (black solid line). Subsequent reversals of the magnetic field direction leads to the formation of the $2\pi$-skyrmion (b) and the $3\pi$-skyrmion (c). The oscillation of the two in-plane rings of the $2\pi$-skyrmion was fitted by Eq.~(\ref{eqn:2}) and subsequently the offset and the additional exponential decay was subtracted in the inset of (b) (black and red solid lines).}
		\label{graphic:fig5}
\end{figure}

We performed LLG simulations starting with a ferromagnetic state at a fixed field value $B$, relaxed to the equilibrium state where the spins at the edge become slightly tilted due to the Dzyaloshinskii--Moriya interaction. At $t=0$\,ps, we reverse the direction of the external magnetic field. As a response to this initial nonequilibrium condition, the spins at the edge of the sample reverse their direction as a function of time initiating the creation of a $1\pi$-skyrmion, by forming a ring where the directions of the spins  lie within the plane of the system. Figure~\ref{graphic:fig5}(a) shows the radius of this ring $R$ as a function of time for the simulation parameters $r=60\,a$ and $B=1\,$T. The radius oscillates and stabilizes as the system converges to the stable skyrmion state after about $100\,$ps. We found that the damped oscillation after $35\,$ps, shown in the inset of Fig.~\ref{graphic:fig5}(a), can be well described by the function 
\begin{align}
R(t) = R_0 + A \exp \left(-\lambda t\right)\sin\left(\omega\left(t-t_c\right)\right),\label{eqn:1}
\end{align}
with the angular frequency $\omega = 272\cdot 10^{9}\,$s$^{-1}$ $($f$ = 43\,$GHz$)$, the decay parameter $\lambda= 181\cdot 10^{9}\,$s$^{-1}$, the equilibrium skyrmion radius $R_0 = 7.2\,a$, the amplitude A and time shift $t_c$. The oscillation frequency $f$, also given in Table~\ref{table1}, is most likely connected to the breathing mode of the skyrmion\cite{Schutte2014a}. This switching process is shown in Supplemental Video 1\cite{supp}.




\begin{table}
\centering
\begin{ruledtabular}
\begin{tabular}{lrr}
  & \multicolumn{1}{c}{$f=\omega/2\pi$ [GHz]} & \multicolumn{1}{c}{$\lambda$ [$10^{9}\,\textrm{s}^{-1}$]} \\ \hline
1$\pi$ & $43.29\pm 0.14$   & $181.49\pm 0.66$    \\
2$\pi$  inner ring  & $88.81\pm 0.47$        & $198.41\pm 3.54$    \\
2$\pi$  outer ring & $89.29\pm 0.80$        & $241.55\pm 4.08$    \\
3$\pi$ inner ring & $38.20\pm 2.01$        & $256.41\pm 32.22$   \\
3$\pi$ middle ring & $91.41\pm 0.33$        & $195.69\pm 1.53$    \\
3$\pi$ outer ring & $101.01\pm 2.44$        & $277.78\pm 9.26$   \\
\end{tabular}
\end{ruledtabular}
\caption{Oscillation frequencies $f$ and decay parameters $\lambda$ of the in-plane rings of $k\pi$-skyrmionic structures after switching the field direction.\label{table1}}
\end{table}

After the skyrmion state is fully converged, we instantaneously reverse the direction of the external magnetic field once more. Again the spins at the edge of the sample reverse their direction as a function of time as a response to the new nonequilibrium condition. Hence, an additional ring emerges after about $4\,$ps, in which the directions of the spins lie within the sample plane, the radius of which is denoted by a red line in Fig.~\ref{graphic:fig5}(b). The radii of the in-plane rings oscillate with a phase shift of about $\pi$ with respect to each other as a function of time and shrink in size as the system converges to the stable $2\pi$-skyrmion. In contrast to the $1\pi$-skyrmion, here a transient decrease of the whole structure in size has to be taken into account besides the damped oscillation. Hence, we fitted the radii in the time interval $22\,\mathrm{ps}\le t\le 45\,\mathrm{ps}$ with the adapted function
\begin{align}
R(t) = R_0 + B\exp(-\mu t) + A \exp \left(-\lambda t\right)\sin\left(\omega\left(t-t_c\right)\right),\label{eqn:2}
\end{align}
which contains an additional exponential decay. The result of the fitting process is shown in the inset of Fig.~\ref{graphic:fig5}(b). The solid lines depict the oscillatory part of the fit curves excluding the exponential decay and the constant offset for better visibility. The oscillation frequencies and decay parameters of the $2\pi$ skyrmion obtained with Eq.~(\ref{eqn:2}) are displayed in Table~\ref{table1}. One can observe that the determined oscillation frequencies of the inner and outer rings are very close to each other, ensuring a phase shift of about $\pi$ between the rings over an extended time period. This switching process is displayed in Supplemental Video 2\cite{supp}.

Again, we reverse the direction of the external magnetic field after the $2\pi$-skyrmion state is fully converged, and subsequently the spins at the edge reverse their orientation as a function of time forming a third in-plane ring, see Fig.~\ref{graphic:fig5}(c). Here we observe that the outer two in-plane rings oscillate with a phase shift of about $\pi$ with approximately the same frequency as identified in Fig.~\ref{graphic:fig5}(b), the inner ring oscillates with a frequency similar to that of the $1\pi$-skyrmion in Fig.~\ref{graphic:fig5}(a), and the whole structure converges to the stable $3\pi$-skyrmion. The oscillation frequencies and the decay parameters can again be found in Table~\ref{table1}. This switching process is illustrated in Supplemental Video 3\cite{supp}.

The simulations were performed by using the damping parameter $\alpha = 0.3$, similar values to which have been determined experimentally in systems displaying noncollinear magnetic structures\cite{Metaxas2007,Beg2017}. In order to check the robustness of this process with respect to changes in the simulation parameters, we performed the calculations for different disk sizes $r\in \{50\,a,70\,a\}$ and magnetic fields $B\in\{0.9\,\mathrm{T},1.1\,\mathrm{T}\}$ as indicated by the corner points of the striped region in Fig.~\ref{graphic:fig4}. We confirmed that the process is stable against weak thermal fluctuations by also performing simulations at $T=5$\,K.

This demonstrates a robust way of inducing a transition from a $k\pi$- to a $(k+1)\pi$-skyrmion state by a nonadiabatic reversal of the direction of the external magnetic field. One possibility for inducing a transition in the reverse direction is increasing the magnetic field above the corresponding critical field at the given disk radius, which may be somewhat delicate for the $2\pi$- and $3\pi$-skyrmions since their critical field values are located very close to each other.

\section{Conclusion}
We investigated both static equilibrium and dynamic properties of $k\pi$-skyrmions with $k\in \{1,2,3\}$ in a classical lattice spin model describing the Pd/Fe/Ir(111) ultrathin film system. We found that $2\pi$- and $3\pi$- skyrmions may be stabilized in the system, but they collapse at significantly lower field values than $1\pi$-skyrmions. By relying on the GNEB method, we determined minimum energy paths between the ferromagnetic state and the $k\pi$-skyrmions, and calculated the energy barriers. We investigated the transition states obtained from the GNEB method to define the minimal size of $k\pi$-skyrmions as a function of magnetic field. Finally, we proposed a dynamical method for crossing the energy barrier from the $k\pi$-skyrmion to the $(k+1)\pi$-skyrmion induced by a nonadiabatic reversion of the direction of the magnetic field in disk-shaped samples of finite size, and investigated the decaying oscillations in the skyrmion size. Our results may stimulate further experimental and theoretical investigations of $k\pi$-skyrmions in ultrathin magnetic films.

\begin{acknowledgments}

Financial support by the Deutsche Forschungsgemeinschaft via SFB 668, by the European Union via the Horizon 2020 research and innovation program under Grant Agreement No. 665095 (MAGicSky), and by the Alexander von Humboldt foundation is gratefully acknowledged.
\end{acknowledgments}

\appendix

\section{Verification of lattice interaction parameters}\label{app}

\begin{figure}[tb]
\includegraphics[width=\columnwidth]{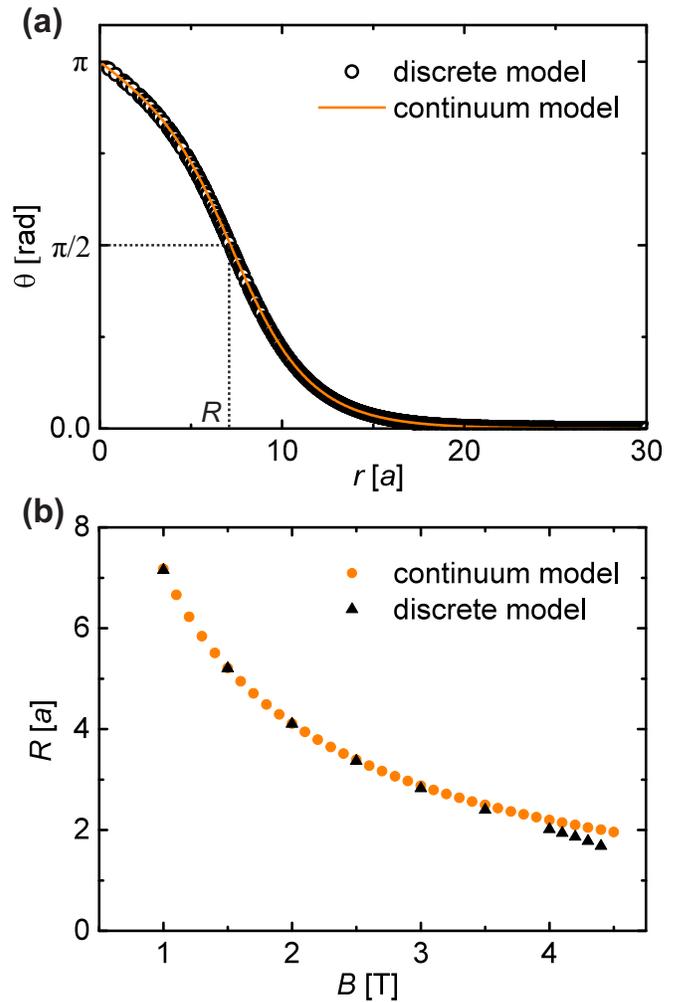}
		\caption{(a) Comparison of the $1\pi$-skyrmion profile as derived in the continuum model and in the discrete model at the magnetic field $B=1\,$T. (b) The radius of the $1\pi$-skyrmion as a function of the magnetic field from the continuum model and the discrete model. A significant deviation between the two descriptions only becomes apparent for large magnetic fields.}
		\label{graphic:app1}
\end{figure}

In Ref.~\cite{Romming2015}, the continuum material parameters $\mathcal{A} = 2\,$pJ/m spin stiffness, $\mathcal{D} = 3.9\,$mJ/m$^2$ Dzyaloshinskii--Moriya interaction, $\mathcal{K} = 2.5\,$MJ/m$^3$ anisotropy term and $\mathcal{M}_s = 1.1\,$MA/m saturation magnetization were determined experimentally for the Pd/Fe/Ir(111) system. We transformed these to atomic interaction parameters on the two-dimensional hexagonal lattice by considering the values $\mu=3\,\mu_{\mathrm{B}}$ and $a=2.71\,\mathrm{nm}$ obtained from density functional theory calculations\cite{Dupe2014} and using the relations $\mathcal{A}/\mathcal{M}_s=3a^{2}J/4\mu$, $\mathcal{D}/\mathcal{M}_s=3D/2\mu$, and $\mathcal{K}/\mathcal{M}_s=K/\mu$. In order to verify the consistency of the two descriptions, we compared the profile of a $\pi$-skyrmion obtained from the solution of the Euler--Lagrange equations\cite{Bogdanov1994} for the continuum model and from LLG simulations performed for the discrete model, respectively. These profiles are displayed in Fig.~\ref{graphic:app1}(a) at $B = 1\,$T, while Fig.~\ref{graphic:app1}(b) presents the radius $R$ (as defined in the main text) of the $\pi$-skyrmion as a function of the magnetic field. We find a very good agreement between the two models, with significant deviations only observable at small skyrmion sizes where the angle between neighboring spins becomes large in the lattice model.


\bibliography{literature}

\end{document}